# BOHR SOMMERFELD QUANTISATION : COMbining PERTURBATION THEORY and asymptotic analysis

Shayak Bhattacharjee

Department of Physics
Indian Institute of Technology Kanpur
Kanpur 208016,

India

D. S. Ray

Department of Physical Chemistry
Indian Association for the Cultivation of Science
Jadavpur
Kolkata 700032
India

and

J. K. Bhattacharjee

S. N. Bose National Centre for Basic Sciences
Salt Lake
Kolkata 700098
India.

ABSTRACT

We use the Bohr Sommerfeld quantization rule along with a perturbative evaluation of the action intergral to find exact energy levels for the Pöschl-Teller potential (both hyperbolic and trigonometric forms), the Morse potential, and the Rosen Morse potential. Combining perturbation theory with the simplest asymptotic evaluation of the action integral allows us to obtain all the energy levels of the Lennard-Jones potential with an accuracy greater than

0.1 per cent and serves to confirm that the perturbation results for Pöschl-Teller and Morse Potentials are exact.

# I. INTRODUCTION :

It is quite common in quantum mechanics texts[1] to obtain the quantised energy levels of certain canonical systems from the action angle quantization which is also the Bohr Sommerfeld quantization condition[2]. This quantization scheme is successful in one dimension when the classical system shows a unique periodic orbit for a given energy (an example of an exception is the double well potential which has two possible periodic orbits for energies which is below the unstable maximum of the potential and the splitting of energy levels does not follow from the usual Bohr Sommerfeld Condition[3]). Quantization of the action integral along the periodic orbit leads to the condition ($p$ is the momentum) :

$$J(E) = \oint p(x)dx = nh \qquad (1.1)$$

where n = 1,2,… . Using the WKB method it can be argued[4] that $n$ should really be $n + \alpha$, where $\alpha$ is a constant which for smooth potentials in one dimension turns out to be $1/2$. Then $n = 0$ can be included in the range of values of $n$. The potentials usually studied using this technique are the infinite square well, simple harmonic oscillator and the general class of oscillators with the potential $V(x) = \alpha |x|^k$, where $k$ is an integer. It was shown by Robinett[5] that perturbation theory can be carried out on Eq. (1.1) and can yield almost trivially the correspondence principle limit of first order perturbation theory. In spite of its simplicity the Bohr Sommerfeld technique remains a popular area of investigation[6,7]. Recent works deal with a two dimensional electron gas in a magnetic field[8] and issues related to 3-branes[9].

Our primary observation is that any potential $V(x)$ with two turning points which also possess a minimum can be expanded about the same. If the minimum is at $x=x_0$ and $V(x_0) = V_0$ then the expansion will acquire the form

$$V(x) = V_0 + \frac{1}{2}V''(x_0)(x-x_0)^2 + \frac{1}{3!}V'''(x_0)(x-x_0)^3 + \frac{1}{4!}V''''(x_0)(x-x_0)^4 + \ldots$$

$$= V_0 + \frac{1}{2}m\omega^2(x-x_0)^2 + \frac{\alpha}{3}(x-x_0)^3 + \frac{\beta}{4}(x-x_0)^4 + \ldots$$

where ω, α and β are easily identified.
With V(x) expanded as above it will be our aim to evaluate the action $J(E) = \oint p dx$ which equals $\oint \sqrt{2m(E-V(x))}dx$, perturbatively about the quadratic term and express it as a power series in α, β etc. Using equation 1.1 we can now obtain $E$ as a power series in α, β …

We will see that for quite a few potentials the coefficients α, β etc conspire to cause the perturbation theory to terminate and we get an exact answer.

Here we will show that using perturbation theory to go with Eq. (1.1), we can include nontrivial potentials[6] like the Pöschl-Teller potential, its trigonometric variant, the Morse potential and even the Rosen–Morse potential in the ambit of Eq. (1.1). On top of that if we include a simple asymptotic analysis, we can also treat potentials of the Lennard Jones kind which is the potential $V(x) = V_0 \left[ \left( \frac{a}{x} \right)^{12} - \left( \frac{a}{x} \right)^6 \right]$, where $V_o$ and $a$ are constants. It is interesting to note that the Lennard-Jones potential keeps drawing a fair amount of attention[7-12] because of its relevance as an intermolecular interaction potential. We manage to provide a decent account of the energy levels by a combination of perturbative and asymptotic calculations. In addition the asymptotic analysis helps confirm that the perturbation theory did give the right answer. We set up the general perturbation theory for Eq. (1.1) in Sec. II treat the Pöschl-Teller and Morse potentials in Sec. III, and the Lennard Jones variety in Sec. IV.

## II. THE PERTURBATION THEORY

In this section we evaluate the action integral $J$ for the basic anharmonic oscillator having the potential $V(x) = \frac{1}{2} m\omega^2 x^2 + \frac{\alpha}{3} x^3 + \frac{\beta}{4} x^4$ using perturbation theory. This requires a perturbative evaluation of the integral

$$J = 2\sqrt{2m} \int_{a_l}^{a_r} \sqrt{E - \frac{1}{2} m\omega^2 x^2 - \frac{\alpha x^3}{3} - \frac{\beta x^4}{4}} \qquad \ldots\ldots (2.1)$$

Where $a_l$ and $a_r$ are the left and right turning points respectively. Since the oscillations are about the origin $a_l < 0$ and $a_r > 0$. If the amplitude of motion is $a$ (i.e. the kinetic energy vanishes at $a = 0$), then

$$E = \frac{1}{2} m\omega^2 a^2 + \frac{\alpha a^3}{3} + \frac{\beta a^4}{4} \qquad \ldots\ldots (2.2)$$

The left and right turning points are the negative and positive values of 'a' that satisfy Eq. (2.2). Our first task is a perturbative evaluation of $a_l$ and $a_r$. To this end we expand

$$a_{l,r} = a_0 + \alpha a_1 + \alpha^2 a_2 + \beta a_1' \qquad \ldots\ldots (2.3)$$

Inserting in Eq. (2.2)

$$E = \frac{1}{2}m\omega^2 \left[ a_0^2 + 2\alpha a_1 a_0 + 2a_0 a_2 \alpha^2 + a_1^2 \alpha^2 + 2a_0 a_1' \beta + \ldots \right]$$
$$+ \frac{\alpha}{3}\left[ a_0^3 + 3\alpha a_0^2 a_1 \right] + \frac{\beta}{4} a_0^4 + \ldots\ldots\ldots\ldots\ldots$$

$$= \frac{1}{2}m\omega^2 a_0^2 + \alpha[m\omega^2 a_0 a_1 + \frac{a_0^3}{3}] + \alpha^2[m\omega^2 a_0 a_2 + \frac{1}{2}m\omega^2 a_1^2 + a_0^2 a_1] + \beta[m\omega^2 a_0 a_1' + \frac{a_0^4}{4}] + \ldots$$
(2.4)

Equating identical powers of α and β from either side of Eq. (2.4)

$$a_0^2 = \frac{2E}{m\omega^2} \qquad \ldots\ldots (2.5)$$

$$a_1 = -\frac{a_0^2}{3m\omega^2} \qquad \ldots\ldots (2.6)$$

$$a_0 a_2 = \frac{5}{18}\left(\frac{a_0^2}{m\omega^2}\right)^2 \qquad \ldots\ldots (2.7)$$

$$a_1' = -\frac{a_0^3}{4m\omega^2} \qquad \ldots\ldots (2.8)$$

There are two roots of $a_0$ from Eq. (2.5), the positive corresponds to $a_r$ and the negative to $a_l$. Accordingly, are can write down the perturbation expression for the two turning points as

$$a_r = |a_0| - \frac{\alpha a_0^2}{3m\omega^2} + \frac{5}{18}\alpha^2 \left(\frac{a_0^2}{m\omega^2}\right)^2 \frac{1}{|a_0|} - \beta \frac{|a_0|^3}{4m\omega^2} \qquad \ldots\ldots (2.9)$$

$$a_l = -|a_0| - \frac{\alpha a_0^2}{3m\omega^2} - \frac{5}{18}\alpha^2 \left(\frac{a_0^2}{m\omega^2}\right)^2 \frac{1}{|a_0|} + \beta \frac{|a_0|^3}{4m\omega^2} \qquad \ldots\ldots (2.10)$$

Having found the turning points, we can now evaluate J in perturbation theory. Accordingly, Eq (2.1) is written as

$$J = 2\sqrt{2m} \int_{a_l}^{a_r} \sqrt{\frac{1}{2}m\omega^2\left(a^2 - x^2\right) + \frac{\alpha}{3}\left(a^3 - x^3\right) + \frac{\beta}{4}\left(a^4 - x^4\right)}$$

$$= 2m\omega \int_{a_l}^{a_r} \sqrt{a^2 - x^2}\left[1 + \frac{\alpha}{3m\omega^2}\frac{a^3 - x^3}{a^2 - x^2} - \frac{1}{18}\left(\frac{\alpha}{m\omega^2}\right)^2\left(\frac{a^3 - x^3}{a^2 - x^2}\right)^2 + \frac{\beta}{4m\omega^2}\left(a^2 + x^2\right) + \ldots\ldots\right]dx$$
(2.11)

Now,

$$\int_{a_l}^{a_r} \sqrt{a^2 - x^2}\,dx = \int_{a_l}^{0} \sqrt{a_l^2 - x^2}\,dx + \int_{0}^{a_r} \sqrt{a_r^2 - x^2}\,dx$$

$$= \frac{\pi}{4}\left(a_l^2 + a_r^2\right)$$

$$= \frac{\pi}{4}\left[2a_0^2 + \frac{4}{3}\alpha^2\left(\frac{a_0^2}{m\omega^2}\right)^2 - \beta a_0^4 + \ldots\ldots\ldots\right]$$

……………….. (2.12)

Similarly,

$$\int_{a_l}^{a_r} \frac{a^3 - x^3}{\sqrt{a^2 - x^2}} = \frac{\alpha}{3m\omega^2}\left(\frac{\pi}{2} - \frac{2}{3}\right)\left(a_r^3 - |a_l|^3\right)$$

$$= -\frac{2}{3}\alpha^2\left(\frac{a_0^2}{m\omega^2}\right)^2\left(\frac{\pi}{2} - \frac{2}{3}\right) + \ldots\ldots$$ ……………….. (2.13)

Working to $O(\alpha^2)$ and $O(\beta)$ implies that the remaining integrals in Eq. (2.11) can be evaluated from $-|a_0|$ to $|a_0|$ and thus

$$-\frac{1}{18}\left(\frac{\alpha}{m\omega^2}\right)^2 \int_{-|a_0|}^{|a_0|} \frac{(a_0^3 - x^3)^2}{(a_0^2 - x^2)^{3/2}} = -\frac{a_0^4}{9}\left(\frac{\alpha}{m\omega^2}\right)^2\left(4 - \frac{15\pi}{16}\right) \quad \ldots\ldots\ldots\ldots (2.14)$$

and

$$\frac{\beta}{4m\omega^2} \int_{-|a_0|}^{|a_0|} \sqrt{a_0^2 - x^2}\,(a_0^2 + x^2)\,dx = \frac{\beta}{4m\omega^2}\,\frac{5\pi a_0^4}{8} \quad \ldots\ldots\ldots\ldots (2.15)$$

Using Eq. (2.5) and putting together the results from Eq. (2.12) to Eq. (2.15) we have

$$J = \frac{2\pi E}{\omega} + \frac{5\pi\alpha^2}{6m^3\omega^7}E^2 - \frac{3\pi\beta}{4\omega}\frac{E^2}{m^2\omega^4} + \ldots\ldots \quad \ldots\ldots\ldots\ldots (2.16)$$

We now use the quantization condition $J = nh$ and solve for $E$ perturbatively by expanding

$$E = E_0 + \alpha E_1 + \alpha^2 E_2 + \beta E_2' + \ldots\ldots\ldots\ldots \quad \ldots\ldots\ldots\ldots (2.17)$$

Using the same steps as for the turning points, we get

$$E = n\hbar\omega - \frac{5}{12}\frac{\alpha^2}{m^3\omega^4}n^2\hbar^2 + \frac{3}{8}\frac{\beta}{m^2\omega^2}n^2\hbar^2 \quad \ldots(2.18)$$

This is one of the central results we will use later.
Sometimes it may be necessary to explore the situation beyond the quartic term. Anticipating this, we exhibit how this is done for the symmetric anharmonic oscillator where the potential is expressed as

$$V(x) = \frac{1}{2}m\omega^2 x^2 + \frac{\beta}{4}x^4 + \frac{\gamma}{6}x^6 \qquad \ldots\ldots\ldots\ldots (2.19)$$

We now need to carry out our calculation to the second order in β and first order in γ. Having shown the details of the calculation in our previous example, here we only sketch the results. The first step is to find the turning points at an energy E. For the symmetric potential, the turning points will be at ±a with 'a' satisfying

$$E = \frac{1}{2}m\omega^2 a^2 + \frac{\beta}{4}a^4 + \frac{\lambda}{6}a^6 \qquad \ldots\ldots\ldots\ldots (2.20)$$

and having the expansion

$$a = a_0 + \beta a_1 + \beta^2 a_2 + \gamma a_3 + \ldots\ldots\ldots \qquad \ldots\ldots\ldots\ldots (2.21)$$

Exactly as in Eq. (2.5) – (2.8), we obtain

$$a_0^2 = \frac{2E}{m\omega^2}, \quad a_1 = -\frac{a_0^3}{4m\omega^2}, \quad a_2 = \frac{7}{32}\frac{a_0^5}{(m\omega^2)^2}, \quad a_3 = -\frac{a_0^5}{6m\omega^2} \qquad \ldots\ldots\ldots\ldots (2.22)$$

The action J can now be evaluated as

$$J = 4\int_0^a \sqrt{2m\left[\frac{1}{2}m\omega^2(a^2 - x^2) + \frac{\beta}{4}(a^4 - x^4) + \frac{\gamma}{6}(a^6 - x^6)\right]}dx \qquad \ldots(2.23)$$

Evaluating the integrals, we are led to

$$J = \pi m a\omega^2\left[1 + \frac{5}{16}\frac{\beta a^2}{m\omega^2} - \frac{13}{256}\beta^2\left(\frac{a^2}{m\omega^2}\right)^2 + \frac{11}{48}\gamma\left(\frac{a^2}{m\omega^2}\right)^2 m\omega^2 + \ldots\ldots\right] \qquad (2.24)$$

Introducing the expansion of 'a' [Eq. (2.21)] and the result shown in Eq. (2.22), we find

$$J = \frac{2\pi}{\omega}\left[E - \frac{3\beta}{8(m\omega^2)^2}E^2 + \frac{35}{64}\beta^2\frac{E^3}{(m\omega^2)^4} - \frac{5}{12}\gamma\frac{E^3}{(m\omega^2)^3} + \ldots\ldots\right] \qquad \ldots\ldots\ldots\ldots (2.25)$$

Expanding the energy E as

$$E = E_0 + \beta E_1 + \beta^2 E_2 + \gamma E_3 + \ldots\ldots\ldots\ldots\ldots \qquad \ldots\ldots\ldots\ldots (2.26)$$

and setting $J = nh$ with J as in Eq. (2.25),

$$E = E_0 + \frac{3}{8}\beta\frac{E_0^2}{(m\omega^2)^2} - \frac{17}{64}\beta^2\frac{E_0^3}{(m\omega^2)^4} + \frac{5\gamma}{12}\frac{E_0^3}{(m\omega^2)^3} \qquad \ldots\ldots\ldots\ldots (2.27)$$

With $E_0 = n\hbar\omega$. This is the analogue of Eq. (2.18) for the symmetric oscillator carried to one higher order in $E_0$. In section III we will show how equations (2.18) and (2.27) can be used to find the eigenvalues of some well known potentials.

## III. APPLICATIONS

In this section we will consider some exactly solvable potentials which require a knowledge of sophisticated special functions for the determination of eigenvalues and eigenfunctions. It will be seen that using equation (2.18) and (2.27) we will be able to obtain the correct eigenvalues in each case.
We begin with the Pöschl – Teller potential

$$V(x) = -V_0 \operatorname{sech}^2 ax \qquad \ldots (3.1)$$

Which is shown in Fig.1. As explained in Sec. I the first step is to expand V(x) about its minimum x=0 and this yields

$$V(x) = -V_0 \operatorname{sech}^2 ax = -V_0 \left[ 1 - a^2 x^2 + \frac{2}{3} a^4 x^4 - \frac{17}{45} a^6 x^6 + \ldots \right] \qquad \ldots (3.2)$$

(3.1)

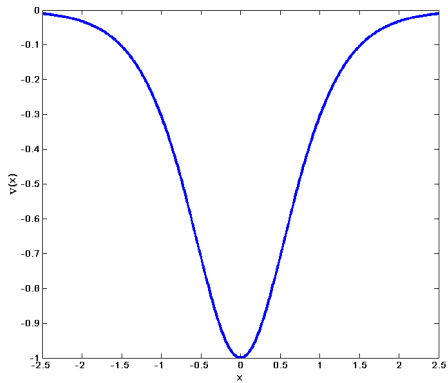

Figure 1  Pöschl Teller Potential

It is clear that bound states can exist only for E < 0. The total energy of this system is

$$E = -V_0 + E' \qquad (3.3)$$

Where E' is the energy associated with the anharmonic oscillator potential

$$V(x) = V_0 \left[ a^2 x^2 - \frac{2}{3} a^4 x^4 + \frac{17}{45} a^6 x^6 - \ldots \right] \qquad (3.4)$$

This leads to the identification of the parameters ω, α, β, γ etc as

$$\omega^2 = \frac{2 V_0 a^2}{m} \;,\; \beta = -\frac{8}{3} V_0 a^4 \;,\; \gamma = \frac{34}{15} a^6 V_0 \qquad (3.5)$$

We first use Eq. (2.18) noting that if the classical motion has two turning points, then the Bohr Sommerfeld quantization condition of Eq. (1.1) reads $J = \oint p dx = \left(n + \frac{1}{2}\right)h$, which simply means that 'n' has to be replaced by $n + \frac{1}{2}$ in all the formulae of Sec. II.

The energy $E'$ associated with the anharmonic oscillator is according to Eq. (2.18), on using Eqs. (3.4)

$$E' = E_0' - \frac{E_0'^2}{4V_0} = -\frac{1}{4V_0}\left(E' - 2V_0\right)^2 + V_0 \qquad (3.6)$$

Where $E_0' = \left(n + \frac{1}{2}\right)\hbar\omega$. To this order, the energy E is consequently,

$$E = -V_0 + E' = -\frac{1}{4V_0}\left[\left(n + \frac{1}{2}\right)\hbar\omega - 2V_0\right]^2$$

$$= -\frac{a^2\hbar^2}{8m}\left[2n + 1 - \sqrt{\frac{8mV_0}{a^2\hbar^2}}\right]^2 \qquad (3.7)$$

The exact answer, obtained from a solution of the Schrodinger equation is known to be

$$E = -\frac{a^2\hbar^2}{8m}\left[2n + 1 - \sqrt{1 + \frac{8mV_0}{a^2\hbar^2}}\right]^2 \qquad (3.8)$$

In the semi-classical limit $8mV_0/a^2\hbar^2 \gg 1$ and it reduces to Eq. (3.7) which as a result is an exact answer.

To make good this claim, we have to demonstrate that at the very least there is no correction to E at the next order i.e. $O(E_0'^3)$ in Eq. (3.6). This is where Eq. (2.27) is useful. Using the constants $\beta$ and $\gamma$ from Eq. (3.4) we see the $O(E_0'^3)$ terms cancel out exactly, leaving us with the answer of Eq. (3.7). We note that there are a finite number of energy values. The maximum allowed value of n in Eq. (3.8) is clearly the integer closest to and lower than $\frac{1}{2}\left[\sqrt{\frac{8mV_0}{a^2\hbar^2}} - 1\right]$. At this point we can evaluate the integral for J directly at $E = 0$ with the exact form $V = -V_0 \operatorname{sech}^2 ax$ of $V(x)$. Equating this value $J_0$ of J to $\left(n_0 + \frac{1}{2}\right)h$, we obtain directly the exact value of $n_0$. The fact that this $n_0$ agrees with that obtained from Eq. (3.7) by simply setting $E = 0$ on the left hand side is a clear indication of the fact that Eq. (3.7) is exact.

For the trigonometric version of the Pöschl – Teller potential, one has

$V(x) = V_0 \sec^2 ax$

$$= V_0 \left[1 + a^2 x^2 + \frac{2}{3} a^4 x^4 + \ldots\ldots\ldots\right] \qquad (3.9)$$

It has a minimum of $V_0$ at x = 0 and rises to infinity at $x = \pm \dfrac{\pi}{2a}$. It is a potential where one makes a transition from a harmonic potential (at x approximately 0) to a confinement by an infinite well (at x approximately $\pm\pi/2a$). Once again we can write the total energy as E = $V_0$ + E` where E` is the energy coming from the anharmonic oscillator $V(x) = \dfrac{1}{2}m\omega^2 x^2 + \dfrac{\beta}{4} x^4$

In the above, $\omega^2 = \dfrac{2V_0 a^2}{m}$ and $\beta = \dfrac{8}{3} V_0 a^4$. Using Eq. (2.18), we immediately see

$$E = V_0 + \left(n + \frac{1}{2}\right)\hbar\omega + \frac{\left[\left(n + \frac{1}{2}\right)\hbar\omega\right]^2}{4 V_0}$$

$$= \left(\sqrt{V_0} + \frac{\left(n + \frac{1}{2}\right)\hbar\omega}{2\sqrt{V_0}}\right)^2 \qquad (3.10)$$

As expected this is the exact result and the $O\left[\left(n + \dfrac{1}{2}\right)^3 \hbar^3 \omega^3\right]$ contribution is zero. It is apparent from Eq. (3.10) that for $\dfrac{\left(n + \dfrac{1}{2}\right)\hbar\omega}{V_0} \ll 1$, the eigenvalues are simple harmonic oscillator like (linear in n), while in the other limit they are infinite well like (quadratic in n).

We now turn to the Morse Oscillator, where

$$V(x) = V_0 \left(e^{-2ax} - 2e^{-ax}\right) = -V_0 + V_0 \left(1 - e^{-ax}\right)^2 \qquad (3.11)$$

This potential is shown in Fig. 2. Bound states are expected for E<0.

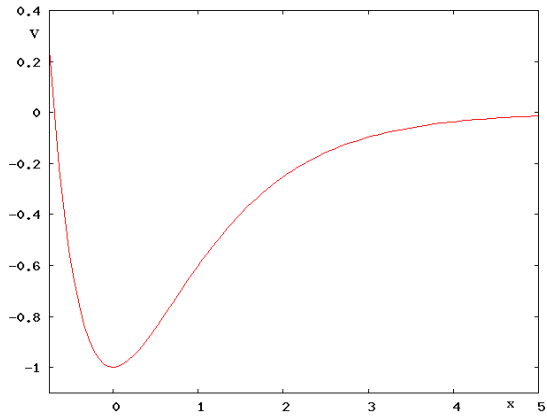

Figure 2       Morse Potential

The minimum of $V(x)$ is at $x = 0$ and as described before, the anharmonic oscillator part of the potential can be found by expanding the right hand side

$$\frac{1}{2}m\omega^2 x^2 + \frac{\alpha}{3}x^3 + \frac{\beta}{4}x^4 + \ldots\ldots \text{with } \omega^2 = \frac{2V_0 a^2}{m}, \ \alpha = -3V_0 a^3, \beta = \frac{7}{3}V_0 a^4 \quad (3.12)$$

Using Eq. (2.18), we find the energy E' of the oscillator as

$$E' = \left(n + \frac{1}{2}\right)\hbar a \sqrt{\frac{2V_0}{m}} - \frac{1}{2}a^2 \frac{\left(n + \frac{1}{2}\right)^2 \hbar^2}{m} \quad (3.13)$$

The total energy E is found to be

$$E = -\left[\sqrt{V_0} - \left(n + \frac{1}{2}\right)\frac{\hbar a}{\sqrt{2m}}\right]^2 \quad (3.14)$$

which is the exact answer for the Morse oscillator.

Once again, to get an idea that this is the exact answer, we need to go to the next order and establish that the contribution is zero. The calculation goes along the lines shown for the Pöschl–Teller potential but is considerably longer since this anharmonic oscillator does not have reflection symmetry. It is however a trivial exercise to carry out an exact evaluation of J for this case at $E = 0$, exactly as in the Pöschl–Teller potential. We write

$$J(E=0) = 2\int_{x_0}^{\infty} \sqrt{2mV_0(2e^{-ax} - e^{-2ax})}\,dx$$ where $e^{-ax_0} = 2$ and the substitution $e^{-ax} = y$ immediately

yields $J(E=0) = \frac{2\pi}{a}\sqrt{2mV_0}$. Equating this to $(n_0 + \tfrac{1}{2})h$ where $n_0$ is the highest quantum level, we get $n_0 + \frac{1}{2} = \sqrt{\frac{2mV_0}{a^2\hbar^2}}$.

The highest quantum level $n_0$ so obtained agrees exactly with the $n_0$ obtained from Eq. (3.14) by setting $E = 0$ on the left hand side. This is a strong confirmation of the fact that Eq. (3.14) is indeed exact.

As a final example we turn to a more complicated potential the Rosen–Morse potential given by

$$V(x) = A^2 + \frac{B^2}{A^2} + 2B \tanh ax - A\left(A + \frac{\alpha\hbar}{\sqrt{2m}}\right) \text{sech}^2 ax \qquad (3.15)$$

The exact eigenvalues of the potential were obtained by Dutt et al[18] using supersymmerty arguments. Writing $\tilde{A} = A + \frac{\alpha\hbar}{\sqrt{2m}}$, we note that the potential has a minimum at $x = x_0$, with

$$\tanh ax_0 = -\frac{B}{A\tilde{A}} \qquad (3.16)$$

Clearly $B < A\tilde{A}$ for there to be a minimum which would allow bound states to exist. Expanding about the minimum

$$V(x) = A^2 + \frac{B^2}{A^2} - A\tilde{A} - \frac{B^2}{A\tilde{A}} + a^2 A\tilde{A} \,\text{sech}^4 ax_0 \,(x - x_0)^2$$

$$+ \frac{a^3}{3}(x - x_0)^3 . 6B \,\text{sech}^4 ax_0 + \frac{a^4}{4}(x - x_0)^4 . \frac{4}{3}\text{sech}^4 ax_0 \left(7\tanh^2 ax_0 - 2\text{sech}^2 ax_0\right) \quad (3.17)$$

Giving rise to the identification

$$m\omega^2 = 2a^2 A\tilde{A} \,\text{sech}^4 ax_0 \qquad \ldots\ldots\ldots\ldots\ldots (3.18)$$

$$\alpha = 6a^3 B \,\text{sech}^4 ax_0$$

$$\beta = \frac{4}{3} A\tilde{A} \,\text{sech}^4 ax_0 (7\tanh^2 ax_0 - 2\text{sech}^2 ax_0)$$

Inserting the above expression for $\omega$, $\alpha$ and $\beta$ in Eq. (2.18) and adding in the constant part of the potential,

$$E = A^2 - A\tilde{A} + \frac{B^2}{A^2}\left(1 - \frac{A}{\tilde{A}}\right) + \left(n + \frac{1}{2}\right)\frac{\hbar\sqrt{2}A}{\sqrt{m}}\left(1 - \frac{B^2}{A^2\tilde{A}^2}\right) - \frac{\left(n + \frac{1}{2}\right)^2 \hbar^2 a^2}{2m}\left(1 + \frac{3B^2}{A^2\tilde{A}^2}\right) + \ldots\ldots$$

$$\ldots\ldots\ldots\ldots(3.19)$$

Since this is a semi-classical approximation, the difference between $A$ and $\tilde{A}$ can be ignored unless essential. Taking note of that, we write Eq. (3.19) as

$$E = A^2 - \left(A - \frac{\left(n+\frac{1}{2}\right)\hbar a}{\sqrt{2m}}\right)^2 - \frac{B^2}{A^4}\left(\left(n+\frac{1}{2}\right)\frac{\hbar\sqrt{2}A}{\sqrt{m}} + \frac{3\left(n+\frac{1}{2}\right)^2 \hbar^2 a^2}{2m}\right) + \ldots \quad (3.20)$$

At this point carrying out the calculation to the next order reveals a term proportional to $B^2/A^4$. We also note that the final exact expression for E cannot have any term which is higher that $n^2$ (the free particle limit). This motivates a reorganization of Eq. (3.20) as

$$E = A^2 + \frac{B^2}{A^2} - \left(A - \left(n+\frac{1}{2}\right)\frac{\hbar a}{\sqrt{2m}}\right)^2 - \frac{B^2}{A^2}\left(1 + 2\left(n+\frac{1}{2}\right)\frac{\hbar a}{\sqrt{2m}}\cdot\frac{1}{A} + 3\left(n+\frac{1}{2}\right)^2 \frac{\hbar^2 a^2}{2m}\cdot\frac{1}{A^2} + \ldots\right)$$

$$= A^2 - \left(A - \left(n+\frac{1}{2}\right)\frac{\hbar a}{\sqrt{2m}}\right)^2 + B^2\left[\frac{1}{A^2} - \frac{1}{\left(A - \left(n+\frac{1}{2}\right)\frac{\hbar a}{\sqrt{2m}}\right)^2}\right]$$

………………… (3.21)

which is the exact answer. We emphasize again that the arrival at the form of the final answer involves Eq. (2.18) and some biased pattern recognition.

## IV. THE LENNARD JONES VARIETY OF POTENTIALS

In this section, we focus on the class of potentials

$$V(x) = V_o\left[\left(\frac{a}{x}\right)^{2k} - \left(\frac{a}{x}\right)^{k}\right] \quad \ldots\ldots\ldots (4.1)$$

For k=6, this gives the well known Lennard Jones potential, shown in Fig. 3.

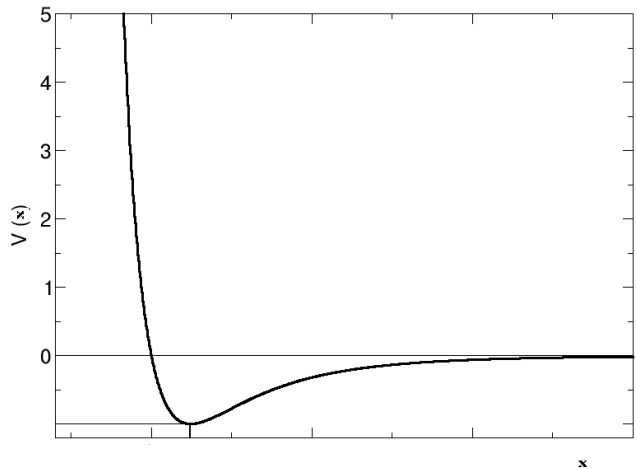

Figure 3   Lennard Jones Potential

Expanded about the minimum at $x_0$ given by

$$2\left(\frac{a}{x_0}\right)^k = 1 \qquad \text{......... (4.2)}$$

the potential takes the form

$$V(x) = -\frac{V_0}{4} + \frac{k^2}{4}V_0\left(\frac{x-x_0}{x_0}\right)^2 - \frac{V_0}{4}k^2(k+1)\left(\frac{x-x_0}{x_0}\right)^3 + V_0\frac{k^2(k+1)(7k+11)}{48}\left(\frac{x-x_0}{x_0}\right)^4 + \ldots$$

......... (4.3)

which is an anharmonic oscillator of the type discussed in Sec. II. Comparing the structure of Eq. 4.3 with the anharmonic potential $V(x) = \frac{1}{2}m\omega^2 x^2 + \frac{\alpha}{3}x^3 + \frac{\beta}{4}x^4$, we have

$$\omega = \frac{k}{\sqrt{2}\, 2^{1/k}}\sqrt{\frac{V_0}{ma^2}}, \quad \alpha = -\frac{3V_0}{4}\frac{k^2(k+1)}{a^2 2^{3/k}}, \quad \beta = \frac{V_0}{12}\frac{k^2(k+1)(7k+11)}{a^4 2^{4/k}} \qquad \text{......... (4.4)}$$

Using Eq. (2.18) and adding in the part $-\frac{V_0}{4}$ to obtain the total energy, we have

$$\frac{4E}{V_0} = -1 + 4\left(n+\frac{1}{2}\right)\frac{\hbar\omega}{V_0} - \frac{2(2k+1)(k+1)}{k^2}\left[\left(n+\frac{1}{2}\right)\frac{\hbar\omega}{V_0}\right]^2 + \ldots \qquad \text{......... (4.5)}$$

We can compare this result with the numerical values given in literature for $k = 6$, the Lennard Jones case. All the numerical results are for $\frac{\hbar\omega}{V_0} = 0.03$ and a comparison between accurate numerical results and our perturbation formula is shown in Table 1.

Table 1

| n | 4E / V₀ from Eq. (4.9) x – (10⁻³) | 4E/V₀ exact. x – (10⁻³) |
|---|---|---|
| 0 | 941 | 941 |
| 1 | 830 | 830 |
| 2 | 728 | 728 |
| 3 | 636 | 634 |
| 4 | 551 | 548 |
| 5 | 477 | 470 |
| 10 | 239 | 186 |

We notice that the accuracy starts deteriorating at n = 5, and the error is already greater than 25 % at n = 10.

We anticipate that our formula will deteriorate progressively as $\frac{4E}{V_0} \to 0$ and accordingly want to look at that limit separately. We want to examine the asymptotic behavior of

$$J = 2\int_{x_1}^{x_2} \sqrt{2m\left\{E - V_0\left[\left(\frac{a}{x}\right)^{2x} - \left(\frac{a}{x}\right)^{k}\right]\right\}}\, dx \qquad (4.6)$$

As $E \to 0$. In Eq. (4.6), $x_1$ and $x_2$ and the turning points of motion with $x_1 < x_2$. Since E is negative, we let $E/V_0 = \epsilon$ (>0) and substituting $y = \left(\frac{a}{x}\right)^k$, we rewrite Eq. (4.6) as.

$$J = \frac{2\sqrt{2mV_0\alpha^2}}{k} \int_{y_1}^{y_2} \sqrt{y - y^2 - |E|}\; \frac{dy}{y^{1+\frac{1}{k}}} \qquad \ldots (4.7)$$

With the two turning points $y_{1,2}$ ($y_1 < y_2$) given by

$$y_{1,2} = \frac{1}{2}\left\{1 \mp \sqrt{1 - 4\varepsilon}\right\} \qquad (4.8)$$

As $E \to 0$, $y_1 \to E$ and $y_2 \to 1$. and the integral of Eq. (4.7) diverges if $k < 2$ and is finite for $k > 2$. Accordingly, we need to treat the two cases separately.

A) <u>k≤2</u> In this case, the integral in Eq. (4.7) diverges for $E \to 0$ and in extracting this divergent behavior from Eq. (4.7) we can drop the $y^2$ term in comparison to $y$. The integral is thus approximated as

$$J \cong 2\frac{\sqrt{2mV_0\alpha^2}}{k} |\varepsilon|^{\frac{1}{2} - \frac{1}{k}} \int_{1}^{1/\varepsilon} \sqrt{z - 1}\; \frac{dz}{z^{1+1/k}} \qquad (4.9)$$

Since $\epsilon \to 0$ we can set the upper limit equal to infinity and thus get

$$J \cong \frac{\sqrt{2\pi V_0 \alpha^2}}{k} \frac{\Gamma\left(\frac{1}{k} - \frac{1}{2}\right)}{\Gamma\left(\frac{1}{k} + 1\right)} \varepsilon^{\frac{1}{2} - \frac{1}{k}} \qquad (4.10)$$

For k<2, this is the dominant contribution to J as $\epsilon \to 0$ Using the Bohr Sommerfeld quatization condition, we have for $\varepsilon \to 0$,

$$\left|\frac{4E}{V_0}\right|^{\frac{1}{2} - \frac{1}{k}} = 2^{2-\frac{1}{k}}\sqrt{\pi}\; \frac{\Gamma\left(1 + \frac{1}{k}\right)}{\Gamma\left(\frac{1}{k} - \frac{1}{2}\right)}\left(n + \frac{1}{2}\right)\frac{\hbar\omega}{V_0} \qquad (4.11)$$

Thus for $\frac{n\hbar\omega}{V_0} \gg 1$, we have Eq. (4.11) and for $\frac{n\hbar\omega}{V_0} \ll 1$, we have Eq. (4.5). The simplest formula connecting both ends is

$$\frac{4E}{V_0} = -\frac{1}{\left\{1 + 2^{2-\frac{1}{k}} \frac{\sqrt{\pi}\,\Gamma\!\left(1+\frac{1}{k}\right)}{\Gamma\!\left(\frac{1}{k}-\frac{1}{2}\right)} \left(n+\frac{1}{2}\right)\frac{\hbar\omega}{V_0}\right\}^{\frac{2k}{2-k}}} \tag{4.12}$$

For k = 1 in particular

$$\frac{4E}{V_0} = -\frac{1}{\left[1 + (2n+1)\frac{\hbar\omega}{V_0}\right]^2} \tag{4.13}$$

which exactly reproduces Eq. (4.5) on expansion. In this case the integral for J can be exactly evaluated and Eq. (4.13) is the exact answer. For other values of k, a simple formula like the one of Eq. (4.12) would not be adequate since on expanding, even the first term in $\left(n+\frac{1}{2}\right)\frac{\hbar\omega}{V_0}$ is not correctly reproduced. We do not concern ourselves with this any more but rather turn our attention to k > 2, which has been popular in the literature.

B) k>2  We return to Eq. (4.7) and examine it for $\varepsilon \to 0$ for k>2. we note that for $\varepsilon = 0$, the integral has a finite value which will clearly be the leading term of J. This finite value, $J_0$, is found as

$$J_0 = 2\frac{\sqrt{2mV_0}\,a^2}{k} \cdot \int_0^1 \sqrt{y(1-y)} \cdot \frac{dy}{y^{1+\frac{1}{k}}}$$

$$= \frac{\sqrt{\pi}}{2^{1/k}} \cdot \frac{\Gamma\!\left(\frac{1}{2}-\frac{1}{k}\right)}{\Gamma\!\left(2-\frac{1}{k}\right)} \frac{V_0}{\omega} \qquad\qquad \text{………………….. (4.14)}$$

This immediately yields the value $n_0$ of n for which $\varepsilon = 0$ since $J_0 = \left(n_o + \frac{1}{2}\right)h$. For the numerically studied case of $\frac{\hbar\omega}{V_0} = 0.03$ and k = 6, This yields $n + \frac{1}{n} = 23.8$ in agreement with earlier results.

Finding the subleading term of J is made difficult by the fact that a simple Taylor expansion of the integrand in Eq. (4.7) leads to divergent integrals. This tells us that the structure of J will be $J_0 + J_1 \varepsilon^\alpha + \ldots\ldots\ldots$, where $\alpha < 1$. Here we will find α but not determine $J_1$. The addition to $J_0$ comes from two sources:

(i) The limits of the integration which are $\varepsilon$ – dependent
(ii) The integrand itself.

Focusing on (i) alone, we find that J should have been written as ($\epsilon \ll 1$)

$$\frac{k}{2} \cdot \frac{1}{\sqrt{2mV_0}\, a^2} \quad J = \int_{\epsilon}^{1-\epsilon} \sqrt{y(1-y)}\; \frac{dy}{y^{1+1/k}}$$

$$= \int_{0}^{1} \sqrt{y(1-y)}\; \frac{dy}{y^{1+\frac{1}{k}}} - \int_{0}^{\epsilon} \sqrt{y(1-y)}\; \frac{dy}{y^{1+\frac{1}{k}}} - \int_{1-\epsilon}^{1} \frac{\sqrt{y(1-y)}}{y^{1+1/k}} dy \quad (4.15)$$

The first term yields $J_0$ while the second has the approximate value $\frac{2k}{k-2}\, \epsilon^{\frac{1}{2}-\frac{1}{k}}$. The third is proportional to $\epsilon^{3/2}$ and hence is much smaller than the second. Analysis of the $\epsilon$ dependence of the integrand also shows a leading behaviour proportional to $\epsilon^{\frac{1}{2}-\frac{1}{k}}$ and hence we conclude that

$$J = J_0 + C\epsilon^{\frac{1}{2}-\frac{1}{k}} + \quad \text{higher order terms} \quad (4.16)$$

for $\varepsilon \ll 1$, where C is a constant that we have not determined.

We now specialize to the Lennard Jones case of $k = 6$ (note that the following results hold for all $k > 2$ and a similar procedure works every time.) The three pieces of information from Eqns. 4.5, 4.14 and 4.16 can be summarized as follows.

i) For $\frac{\hbar\omega}{V_0} \ll 1$, the expansion of Eq. (4.12) gives

$$\frac{4E}{V_0} = -1 + 4(n+\frac{1}{2})\frac{\hbar\omega}{V_0} - \frac{91}{18}\left(n+\frac{1}{2}\right)^2 \left(\frac{\hbar\omega}{V_0}\right)^2 + \ldots\ldots\ldots\ldots \quad (4.17)$$

ii) $E = 0$ at $n = n_0$ such that

$$n_0 + \frac{1}{2} = \frac{1}{2^{5/6}} \cdot \frac{1}{\sqrt{\pi}} \cdot \frac{\Gamma\left(\frac{1}{3}\right)}{\Gamma\left(\frac{11}{6}\right)} \cdot \frac{V_0}{\hbar\omega} \quad (4.18)$$

(iii) Near $n_o$, E vanishes as $(n - n_0)^3$. The specific value of $\frac{\hbar\omega}{V_0} = 0.03$ which has been extensively numerically investigated, we have $n_0 + \frac{1}{2} = 23.8$.

We now need to find a fitting formula that will match the three pieces of information given above. A possible candidate is

$$\frac{4E}{V_0} = -\frac{\left(1 - \dfrac{n + 1/2}{n_0 + 1/2}\right)^3}{1 + \alpha\left(n + \dfrac{1}{2}\right) + \beta\left(n + \dfrac{1}{2}\right)^2} \tag{4.19}$$

The coefficients α and β are to be determined so that the expansion given in Eq. (4.17) is reproduced and this gives
α = -0.00605     β = 2.7 X 10⁻⁵                                                    (4.20)

The resulting values of $4E/V_0$ are shown in Table II and compared with the exact values.

TABLE II

| n | E (Exact Numerical) X – (10⁻⁵) | E (Fitting Formula) X – (10⁻⁵) |
|---|---|---|
| 0 | 94104 | 94113 |
| 1 | 83000 | 83007 |
| 1 | 72764 | 72770 |
| 3 | 63369 | 63373 |
| 4 | 54785 | 54788 |
| 5 | 46982 | 46984 |
| 6 | 39930 | 39932 |
| 7 | 33596 | 33598 |
| 8 | 27947 | 27950 |
| 9 | 22951 | 22954 |
| 10 | 18572 | 18576 |
| 11 | 14775 | 14779 |
| 12 | 11523 | 11526 |
| 13 | 08777 | 08779 |
| 14 | 06498 | 06500 |
| 15 | 04647 | 04647 |
| 16 | 03181 | 03180 |
| 17 | 02059 | 02055 |
| 18 | 01235 | 01231 |
| 19 | 00666 | 00661 |
| 20 | 00305 | 00300 |
| 21 | 00105 | 00102 |
| 22 | 00019 | 00019 |
| 23 | 00000 (to given accuracy) | 00000 (to given accuracy) |

The agreement is to within 0.1%, which is a significant fact for an analysis which is so straightforward.

# CONCLUSION

It is the simplicity of the Bohr Sommerfeld quantization scheme that has kept it popular[19,20] even now. We have shown that this condition can be effectively used for a wide variety of potentials where it has not traditionally been applied. There are two essential ingredients in our calculation. The first is the expansion of the potential about its minimum. This expression gives an anharmonic oscillator and treating the anharmonic terms as perturbation we can find a perturbative form for the energy. The perturbation technique works well if the energy is close to a minimum. There are some cases (Morse potential, Pöschl Teller potential) where the series terminates at some order i.e. the contribution of all succeeding terms vanishes. In such a case one has an exact answer. The fact that the answer is exact can be corroborated by an asymptotic analysis. This works when we have a finite number of bound states. If the maximum possible value of the bound state energy is $E_{max}$ then the action can be evaluated at $E=E_{max}$ and this allows us to get the highest possible $n_{max}$. $n_{max}$ can also be obtained from the perturbation expansion and if the two agree then one can be confident that the expansion has terminated. If the expansion does not terminate then an analysis near $n_{max}$ can be combined with the perturbation expansion in a Pade type interpolation formula to yield an effective expression for the energy. This is what happened here.

\* \* \* \* \*